\documentclass[useAMS,usenatbib]{mn2e}
\usepackage{graphicx,color}
\usepackage{amsmath}
\usepackage[colorlinks=true,citecolor=blue,linkcolor=red,urlcolor=blue]{hyperref}
\usepackage{dcolumn}
\usepackage{multirow}
\usepackage{url}
\usepackage{times}
\newcommand{\s}{$\sim\, $}

\newcommand{\arcs}{$\arcsec$}

\title[Cavities, shocks and a cold front around 3C~320]{Cavities, shocks and a cold front around 3C~320}
\author[Vagshette et.al.]{Nilkanth D. Vagshette$^{1,2}$\thanks{E-mail: nilkanth@prl.res.in}, 
Sachindra Naik$^{2}$\thanks{E-mail: snaik@prl.res.in}, Madhav. K. Patil$^{3}$\thanks{E-mail: 
patil@associates.iucaa.in} \\ \\
$^{1}$ Department of Physics and Electronics, Maharashtra Udayagiri Mahavidyalaya, Udgir - 413 517, India\\
$^{2}$ Physical Research Laboratory, Ahmedabad - 380 009, India \\
$^{3}$School of Physical Sciences, Swami Ramanand Teerth Marathwada University, Nanded - 431 606, India.\\
}

\begin{document}
\pagerange{\pageref{firstpage}--\pageref{lastpage}} \pubyear{2016}
\maketitle
\label{firstpage}

\begin{abstract}
We present results obtained from the analysis of a total of 110~ks {\it Chandra}
observations of 3C~320 FR~II radio galaxy, located at the centre of a cluster of galaxies 
at a redshift $z=0.342$. A pair of X-ray cavities have been detected at an average distance 
of $\sim$38 kpc along the East and West directions with the cavity energy, age and total power 
equal to $\sim$7.7$\times$10$^{59}$ erg, $\sim$7$\times$10$^7$ yr and $\sim$3.5$\times$10$^{44}$ 
erg s$^{-1}$, respectively. The cooling luminosity within the cooling radius of $\sim$100 kpc was 
found to be $L_{cool} \sim8.5\times10^{43}$ erg s$^{-1}$. Comparison of these two estimates 
implies that the cavity power is sufficiently high to balance the radiative loss. A pair of 
weak shocks have also been evidenced at distances of $\sim$47 kpc and $\sim$76 kpc surrounding 
the radio bubbles. Using the observed density jumps of $\sim$1.8 and $\sim$2.1 at shock 
locations along the East and West directions, we estimate the Mach numbers ($\mathcal{M}$) 
to be $\sim$1.6 and $\sim$1.8, respectively. A sharp surface brightness edge was also
detected at relatively larger radius ($\sim$80 kpc) along the South direction. Density 
jump at this surface brightness edge was estimated to be $\sim$1.6 and is probably 
due to the presence of a cold front in this cluster. The far-infrared luminosity yielded 
the star formation rate of 51 M$_{\odot}$ yr$^{-1}$ and is 1/4$^{th}$ of the cooling rate 
($\dot{M}$ $\sim$ 192 M$_{\odot}$ yr$^{-1}$). 

\end{abstract}


\section{Introduction} 

Detection of copious amount  of X-ray emission from hot gas in the intra-cluster 
medium (ICM) suggests that the 45\% \citep{2006MNRAS.372.1496S} and 44\% -- 64\% 
\citep{2017ApJ...843...76A} of cores of clusters appear to cool faster than the Hubble 
time in the local universe and are referred to as the cool core clusters. It is expected 
that the cooling ICM must fall on to the core at a rate up to 1000 M$_{\odot}$ yr$^{-1}$, 
thereby losing its energy in the form of radiation, mostly in the form of X-rays 
\citep{1994ARA&A..32..277F}. However, the observed star formation rates in the cores 
of such clusters are much lower than the cooling rates \citep{2008ApJ...681.1035O} 
leading to the cooling flow problem. Some kind of heating is, therefore, required to 
explain the observed low star formation rate in the cores of such clusters 
\citep{2006MNRAS.373L..65H,2008ApJ...681..151C}. 

Recent understanding is that the Active Galactic Nucleus (AGN) residing at the core 
of the cluster releases the required amount of energy through the feedback and hence 
resolves the  problem \citep{2006MNRAS.373..959D,2006ApJ...652..216R,2007ARA&A..45..117M}. 
Such a feedback by the AGN has been evidenced in several of the clusters, in the form of 
X-ray deficient regions (cavities or bubbles) as the most discernible features of the 
interaction between the AGN and the ICM \citep{2010ApJ...712..883D,2012MNRAS.421..808P,
2013Ap&SS.345..183P,2015Ap&SS.359...61S,2016MNRAS.461.1885V,2017MNRAS.466.2054V}. These 
cavities are believed to be created due to the AGN outburst by injecting the relativistic 
plasma into the ICM \citep{2004ApJ...607..800B,2006ApJ...652..216R}.

The powerful radio jets launched by the central AGNs also induce weak shocks in the ICM. 
Deep observations employing high resolution instruments onboard X-ray observatories such 
as {\it Chandra} and {\it XMM-Newton} have enabled us to detect large scale weak shocks 
around the powerful radio lobes in several of the galaxy clusters \citep{2005ApJ...628..629N,
2005Natur.433...45M,2006MNRAS.366..417F,2007ApJ...665.1057F,2009A&A...495..721S,
2010MNRAS.407.2046M,2011ApJ...732...13G,2012ApJ...749...19K,2014MNRAS.442.3192V,
2016MNRAS.461.1885V}. These shocks are found to be weak with the Mach numbers in 
the range of 1.2 to 1.7, therefore heating due to such weak shocks is significant 
only at smaller radii \citep{2012NJPh...14e5023M}. Apart from shocks, several other 
substructures are also evident in the central region of the cool core clusters. One 
of such features is the cold front which is seen in several of the merging clusters 
\citep{2000ApJ...541..542M,2001ApJ...549L..47V,2001ApJ...551..160V,2002AstL...28..495V}.  
These cold fronts e.g. cool and dense cores of merging subclusters, travel through 
the hot and shocked outer gas halo of the system, causing density and temperature 
discontinuities in their profiles and surface brightness edges in the X-ray 
images \citep{2001ApJ...549L..47V}. Simulation based studies have confirmed 
their origin through the interaction of two subclusters where gas halos of outer 
subclusters are shocked and stopped due to the inner low entropy gas, coupled with 
their host dark matter of sufficiently high density \citep{2005MNRAS.357..801M,
2007PhR...443....1M}. Cold fronts have been reported in several of the cool core 
clusters. \cite{2010A&A...516A..32G} carried out a sample survey of 45 galaxy 
clusters out of which 19 clusters with low redshift ($z <$ 0.2) were found to 
host such cold fronts. Using {\it Chandra} images, \cite{2009ApJ...704.1349O} 
have detected cold fronts in several galaxy clusters at redshift of 0.05 to 0.3. 
There are several other cluster studies that report detection of such cold fronts 
in the core of galaxy clusters \citep{2001ApJ...555..205M,2009ApJ...704.1349O,
2011ApJ...728...27O, 2011MNRAS.415.3520H,2013ApJ...764...82B,2016ApJ...819..113O}. 

3C~320 is a classical FR~II radio galaxy situated at the centre of a cluster 
 \citep{1985PASP...97..932S} at a redshift of $z=0.342$, positioned at RA=15h 
31m 25.38s; DEC=+35d 33m 40.46s \citep{1996AJ....111.1945D}. 3C~320 is a powerful 
radio galaxy and has been studied in several of the radio bands 
\citep{2002AJ....124.1239H,1991ApJ...371..478M,1986A&AS...65..485R,2001MNRAS.321...37S}. 
The detection of X-ray cavities in the cluster  environment as an outcome of 
the {\bf interaction} with radio lobes has been reported by \cite{2013ApJS..206....7M} . 
Using Very Large Array (VLA) X-band observations, \cite{2002AJ....124.1239H} have 
measured the angular size of the lobe (lobe to lobe distance) to be 15.3 arcsec, 
projected bending angle of 7$\,^{\circ}$ and the lobe length asymmetry of 1.2. 
Infrared study also confirmed the presence of dust in 3C~320 \citep{2002A&A...381..389A}. 
However, a thorough and detailed X-ray study of the cluster has not yet been carried out. 
In the present work, a comprehensive imaging and spectral study of 3C~320 FRII radio 
galaxy is attempted by using high spatial and spectral resolution data from {\it Chandra} 
X-ray observatory. 

This paper present results obtained from systematic analysis of two {\it Chandra} 
observations of 3C~320 cluster for a total exposure of 110~ks. This paper is structured 
as: Section-2 describes observations and data analysis techniques, while Section-3 
discusses the results obtained from X-ray imaging, spectral fitting and comparison 
with the radio maps. Section-4 briefly describes the cavity energetics, shocks and 
cold front detections and finally Section-5 summarizes results from the present study. 
Throughout this paper, we adopt a $\Lambda CDM$ cosmology with $\Omega_m$ = 0.3, 
$\Omega_{\Lambda}$=0.7 and $H_0$ = 70 km s$^{-1}$ Mpc$^{-1}$. Using these 
cosmological parameters and the cosmological 
relations\footnote{http://www.astro.ucla.edu/~Ewright/CosmoCalc.html}, 
the angular size distance D$_A$ = 1007 Mpc corresponds to a scale of 4.9 kpc 
arcsec$^{-1}$. All the errors quoted in this paper are at 68$\%$ (1$\sigma$) 
confidence limit, unless otherwise stated.

\begin{figure}
\vbox{
\includegraphics[trim=00 .7cm 00 00,clip=yes,width=8cm,height=6.5cm]{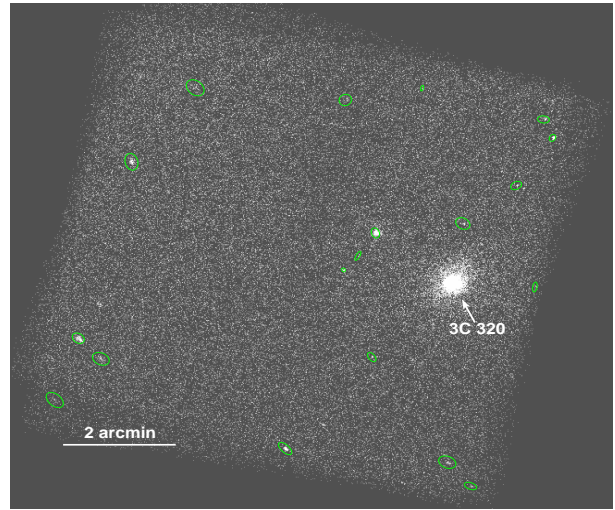}}
\caption{Exposure corrected,  \textit{Chandra} image of 3C~320 galaxy. The green 
ellipses show the point sources detected in the filed of view of S3-chip. In the 
image, North and East are in to the up and left directions, respectively.}
\label{ptsrc} 
\end{figure} 

\begin{figure}
\vbox{
\includegraphics[trim=00 .7cm 00 00,clip=yes,width=9cm,height=7cm]{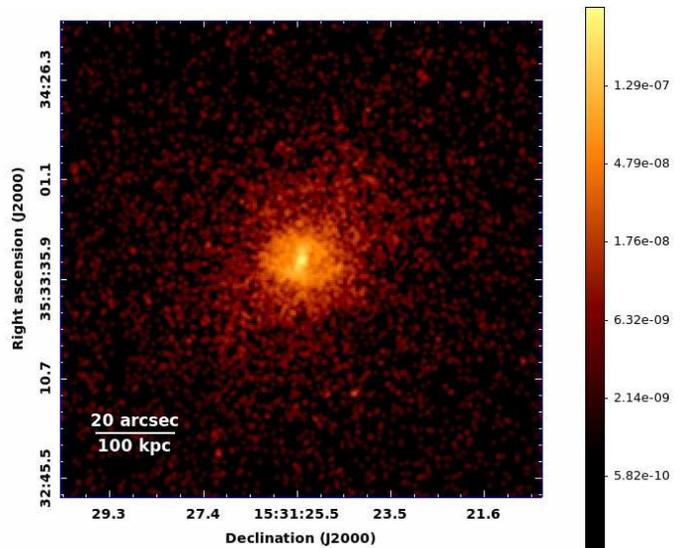}}
\caption{Exposure corrected, 0.5-3 keV $1'5 \times 1'5$ \textit{Chandra} image of 
3C~320 galaxy. }
\label{exp} 
\end{figure} 

\section{Observations and Data Reduction}

The FR~II galaxy 3C~320 was observed with the {\it Chandra} X-ray observatory at two 
epochs; (i) on 2014 May 11 (ObsID 16130) for a total exposure of $\sim$60 ks and (ii) on 
2014 June 12 (ObsID 16613) for a total exposure of $\sim$50 ks. Both the observations were 
carried out in very faint mode (VFAINT) with 3C~320 centered on ACIS-S3 back-illuminated 
chip with other ACIS-2368 chips in switched-on mode. For the present study, data from both 
the observations were retrieved from the Chandra Data Archive 
(CDA\footnote{http://cda.harvard.edu/chaser}) and were reprocessed from level-1 
event files using `chandra$\_$repro' task of {\it CIAO} version 4.6 and {\it CALDB} 
version 4.6.2. Background event files were generated after excluding the hot pixels and 
bright sources from the observed event files from ACIS-S3 chip. Light curves in 2.5--7 
keV range were extracted from these background event files by using `dmextract' task of 
{\it CIAO} and the flaring events were removed by using `lc$\_$sigma$\_$clip' task of 
ChIPS plotting package. In both the observations of 3C~320, no significant flaring 
events were identified. Therefore, we used data with a net exposure of 110~ks for 
further analysis.

\begin{figure*}
\includegraphics[width=16cm,height=7cm]{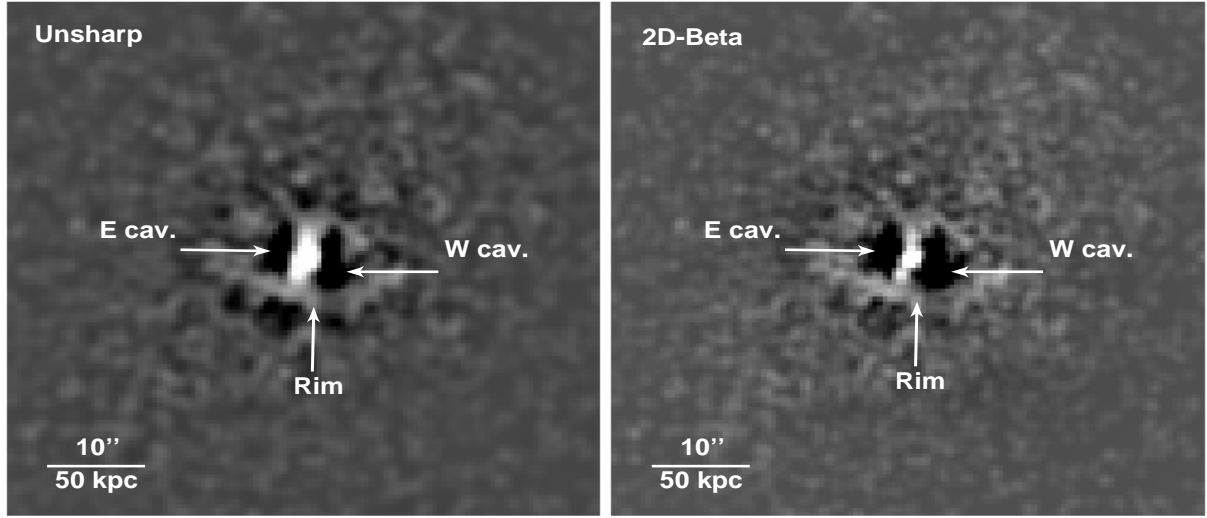} 
\caption{0.5-3 keV residual images of the central 1$\arcmin$ $\times\,$ 1$\arcmin$ 
field of 3C~320: \textit{Left panel} shows the residual image generated by unsharped 
mask imaging technique and \textit{right panel} shows the 2D-$\beta$ model subtracted 
residual image.}
\label{res} 
\end{figure*} 
 
\begin{figure*}
\includegraphics[trim=00 00 00 0.5cm, clip=yes, width=8cm,height=6.5cm]{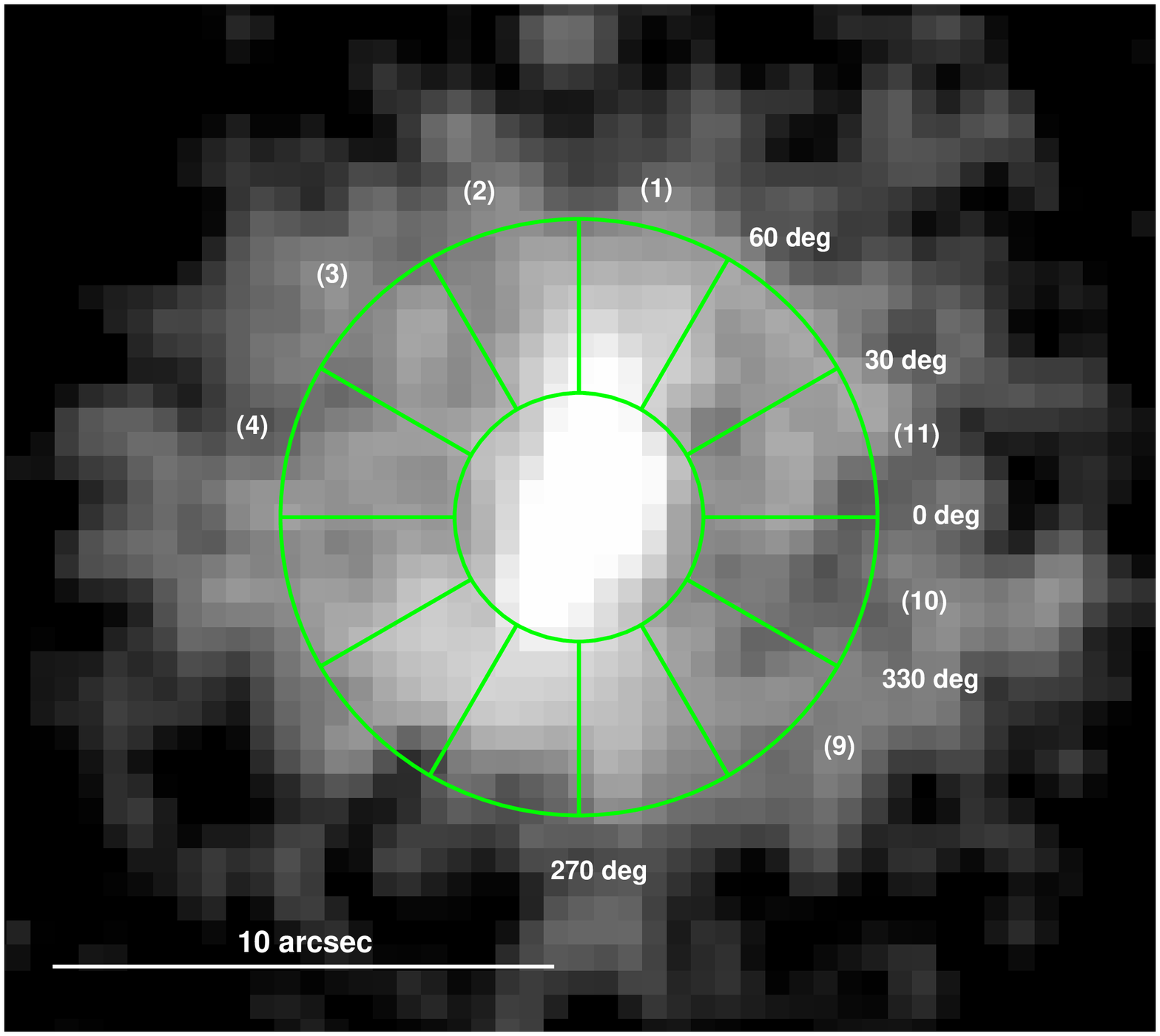} 
\includegraphics[width=8cm,height=7cm]{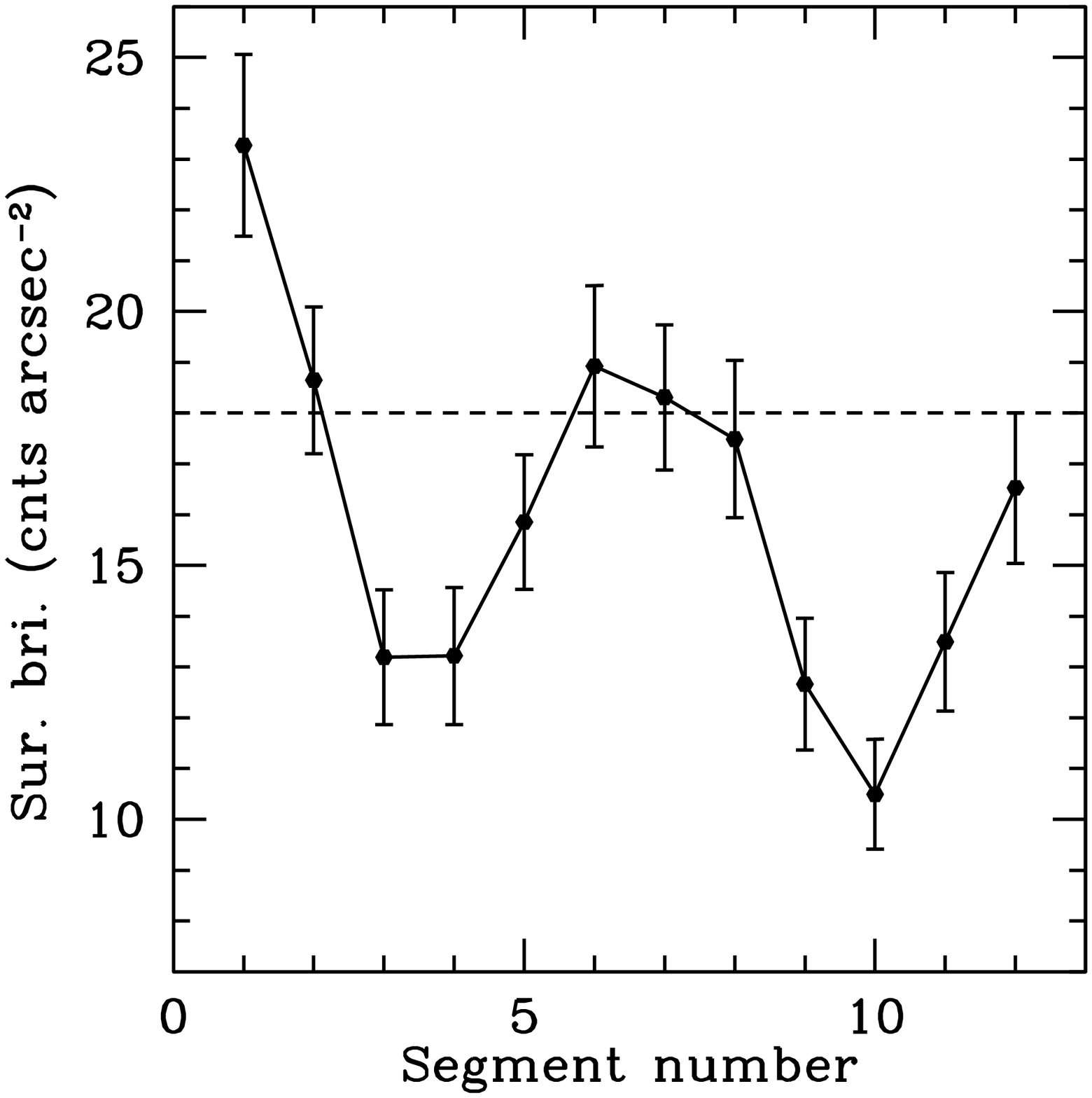} 
\caption{{\it Left panel} shows the count statistics of the sectors from annular 
region of 2-6 arcsec radii from which surface brightness were extracted. {\it Right 
panel} shows the plot of the X-ray counts versus sector numbers. X-ray deficiencies 
along the sectors 3, 4, 9, 10 and 11 depict the X-ray cavities. The horizontal 
dotted line indicate the mean value of X-ray photons.}
\label{seg} 
\end{figure*} 

\section{Results}
\subsection{Imaging analysis}

To improve the signal-to-noise ratio, the cleaned event data from both the observations 
were merged by using `merge$\_$obs\footnote{http://cxc.harvard.edu/ciao/threads/combine/}'. 
In the merged event data, several point sources were identified (Fig.~\ref{ptsrc}) which 
were subsequently removed from the event files by using `wavdetect' task. The resulting 
holes left behind were filled by using `dmfilth' task and the resultant images were used 
for deriving unsharp mask image to delineate the substructures within surface brightness 
distribution of X-ray emission. Exposure corrected, point sources removed and subsequent 
refilled raw image of 3C~320 in 0.5-3 keV range is shown in Fig.~\ref{exp}. 

Identification of peculiar features such as cavities, shocks, etc. in this cluster 
were done by following unsharp and 2D $\beta$-model subtracted residual imaging 
techniques \citep{2009ApJ...705..624D,2010ApJ...712..883D}. In unsharp mask 
imaging technique, a highly Gaussian smoothed image - 7$\sigma$ (1$\sigma$ = 
0.5 arcsec) was subtracted from the lightly smoothed (1$\sigma$) image. In the 
second method, a smooth model image of the cluster emission was generated by 
2D~$\beta$-model fitting by assuming spherical symmetry in the X-ray emission 
which was then subtracted from the original unsmoothed image. The best-fitted 
model yielded the core radius $r_0$ $\sim$ 9.0 arcsec, slope $\alpha$ $\sim$ 1.3 
and amplitude $\sim$ 1.6e-7 counts cm$^{-2}$ arcsec$^{-2}$ s$^{-1}$. The residual 
images generated by using unsharp mask and 2D $\beta$-model subtraction are 
shown in the left and right panels of Fig.~\ref{res}, respectively. In both the 
panels, X-ray cavities along East and West directions of the X-ray centre (dark
shades) of 3C~320, bright circular rims and nuclear clump (excess emission) regions 
are clearly discernible. The inflating cavities that push the infalling gas away are
known to be the cause of the formation of bright rims and nuclear clump (part of the 
rim) gas in the core of 3C~320. These features are very similar to those seen in 
Cygnus~A \citep{2010ApJ...714...37Y}, RBS~797 \citep{2011ApJ...732...71C}, 
MS~0735.6+7421 \citep{2005Natur.433...45M} and 3C~444 \citep{2016MNRAS.461.1885V}. 
To substantiate the depressions in the X-ray emission along the cavity regions, we 
performed X-ray count statistics by extracting the counts from different sectors 
within the annular regions of 2-6 arcsec as shown in the left panel of Fig.~\ref{seg}. 
The right panel of the figure shows the variation of the X-ray counts against the sector 
numbers. In sectors 3 \& 4 (E-cav) and 9, 10 \& 11 (W-cav), we find appreciable decrement 
in the X-ray counts relative to the mean count value which was calculated by using counts 
from the non-deviated regions (sectors 2, 6, 7 and 8) and was found to be \s 18.3 counts 
arcsec$^{-2}$ with a measurement uncertainty of 0.3 counts arcsec$^{-2}$ and standard 
deviation $\sigma \approx$ 0.62 counts arcsec$^{-2}$. The average counts in the regions 
covering the East and West cavity locations were $\sim$13 counts arcsec$^{-2}$ and $\sim$12 
counts arcsec$^{-2}$, with uncertainties of 0.1 counts arcsec$^{-2}$ and 0.9 counts 
arcsec$^{-2}$ and standard deviations of 0.2 counts arcsec$^{-2}$ and 1.6 counts 
arcsec$^{-2}$, respectively. This count statistics delineates more than 3$\sigma$ 
level detection of cavities.

\subsection{Spectral analysis}

To investigate the thermodynamical properties of cluster emission, we performed spectral 
analysis of the data obtained from both the {\it Chandra} observations. As in imaging 
analysis, contributions from the point sources were excluded during the spectral analysis. 
Spectra from each of the observations were extracted independently by using `specextract' 
task of {\it CIAO}. 3C~320 being a moderately redshifted ($z$ = 0.342) cluster, it occupied 
smaller area on the chip. Considering this, we selected local background regions on the 
chip itself, far away from the source emission. Spectra extracted for corresponding 
features from both the observations were simultaneously fitted by using {\it XSPEC} 
version 12.9.0. During spectral fitting, the absorption due to equivalent hydrogen
column density selected within 1$^{\circ}\times1^{\circ}$ field of view was fixed at 
the Galactic value of 1.69$\times$10$^{20}$ cm$^{-2}$ \citep{2005A&A...440..775K}.
 
\begin{figure}
\centering
\includegraphics[width=8.cm,height=13.cm]{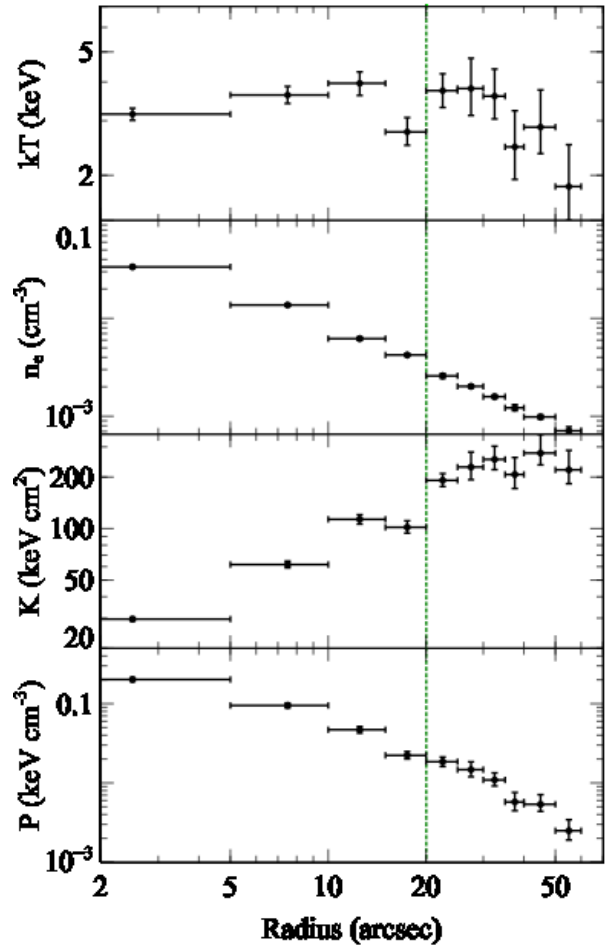} 
\caption{Azimuthally averaged thermodynamical parameters obtained from the spectral 
fitting of data obtained from concentric annular regions from the cluster centre to 
60 arcsec radius. Top to bottom panels show the temperature, electron density, entropy 
and pressure profiles, respectively. The vertical dotted line shows the position 
at which a jump in the temperature profile has been observed.}
\label{therm} 
\end{figure}

To carry out a systematic and thorough understanding of the thermodynamical 
properties in the ICM of the cluster, 0.5--7 keV spectra were extracted from 
concentric circular annuli of 5 arcsec width up to 40 arcsec radius, which 
was then increased to 10 arcsec between 40 to 60 arcsec. These spectra 
were then fitted with a single temperature collisional equilibrium plasma APEC 
model \citep{2001ApJ...556L..91S} allowing the temperature, abundance and 
normalization parameters to vary. Due to poor statistics for a few of the outer 
annuli, beyond 25 arcsrc, the abundance was fixed at 0.5 Z$_{\odot}$ as the 
average abundance of the cluster. For spectral fitting, we used the {\it angr} 
\citep{1989GeCoA..53..197A} abundance table. Electron number density in each 
of the annulus was estimated from APEC normalization which is directly related 
to the emission integral (EI = $\int\, n_e n_H\, dV$) 
\citep{2001ApJ...556L..91S,2010A&ARv..18..127B}. Here, the ratio between 
electron number density ($n_e$) and hydrogen number density ($n_H$) was 
assumed to be 1.2 for solar abundance \citep{2012AdAst2012E...6G}. The radial 
thermodynamical profiles e.g. azimuthally averaged projected temperature, 
electron density, electron entropy ($K=kTn_e^{-2/3}$) and pressure ($\it{p}=nkT$)
derived from this analysis are shown in Fig~\ref{therm}. The temperature profile 
remains almost constant up to 30\arcs~ though it shows a small drop between 15-20\arcs~ 
bin. The temperature in the central 5\arcs~ bin is somewhat lower than the 
temperature in the  10--15\arcs~ bin. The entropy profile exhibits a gradual rise 
from the central lower value while the density and pressure profiles exhibit gradual 
decrease from the centre and are in agreement with the observations in several 
other cool core clusters \citep{2004ApJ...607..800B,2005ApJ...635..894F,
2006ApJ...652..216R,2006MNRAS.373..959D,2007ARA&A..45..117M,2008MNRAS.385..757D,
2016MNRAS.461.1885V,2017A&A...604A.100G,2017ApJ...847...94S,2017ApJ...848...26G}. 
The vertical dotted line in the figure marks the position at which a sharp jump in 
temperature was observed. This temperature jump at $\sim$20 arcsec, far away from 
the cluster centre, indicated the presence of a cold front in the ICM. A detailed 
investigation on the cold front is presented in Section~4.3. Due to relatively 
short exposure of each of the {\it Chandra} observations of the source at a redshift 
of 0.342, it was difficult to obtain the sectoral thermodynamical profiles of the 
cluster emission.

\begin{figure}
\centering
\includegraphics[width=7cm,height=7cm]{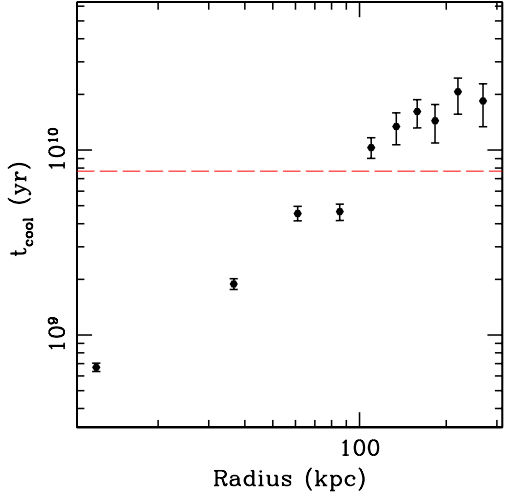} 
\caption{Cooling time profile of the ICM of 3C~320. The dashed line 
indicates the light travel time at $z$=1.}
\label{tcool} 
\end{figure} 

\begin{figure}
\centering
\includegraphics[width=8.cm,height=9.cm,angle=-90]{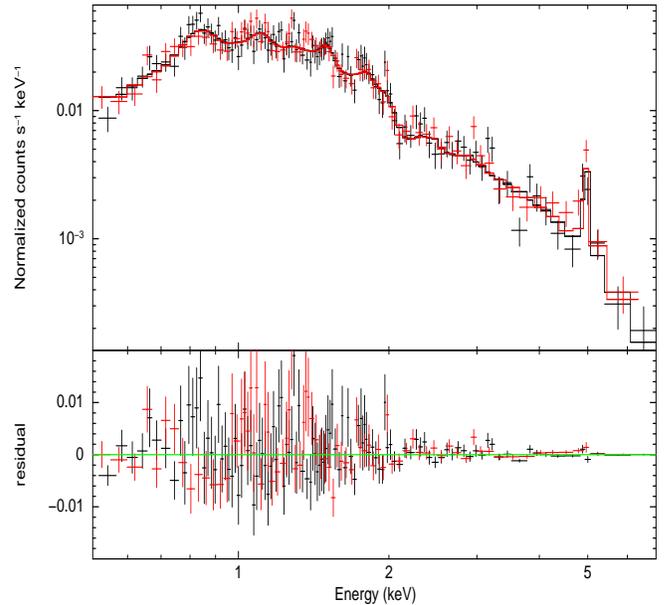} 
\caption{Spectra extracted from within the cooling radius (central 
$\sim$100 kpc) of the cool core cluster 3C~320. The solid line shows the 
best-fit model. Several redshifted emission lines are apparent in the 
spectrum of 3C~320. }
\label{spec} 
\end{figure}

The cooling time for this cluster was derived by using the relation  
\begin{equation}
t_{cool}=\frac{5}{2} \frac{nkT}{n_en_H \Lambda (T,Z)} 
\label{cool}
\end{equation}
where n$_e$, n$_H$, $kT$ and $\Lambda (T,Z)$ represent the electron and 
hydrogen number densities, temperature and the cooling function, respectively 
\citep{2009ApJS..182...12C}. Using this expression, cooling time profile was 
generated and is shown in Fig.~\ref{tcool}. The horizontal dashed line 
in the figure represents the look-back time at 7.7 Gyr \citep{2006ApJ...652..216R}.
The cooling radius R$_{cool}$, the radius at which cooling time corresponds to 7.7 
Gyr, was estimated to be $\sim$100 kpc ($\sim$20 arcsec). Eqn.~\ref{cool} yields 
the cooling time of the gas in the core of this cluster to be $\sim$7 $\times$ 10$^8$ 
yr. Cooling luminosity of the ICM was also estimated by spectral fitting the data 
extracted from a circular region of 100 kpc radius from the core of the cluster 
centre. The 0.5--7 keV spectrum was fitted well with the same collisional equilibrium 
plasma (APEC) code modified with fixed Galactic absorption. The best-fit parameters 
obtained from the fit are -- temperature $kT$ = 3.41$_{-0.09}^{+0.12}$ keV and metal 
abundances $Z$= 0.48$\pm$0.07. The reduced $\chi^2$ for the best-fit model was 1.04 
($\chi^2$/dof = 218.46/209). The cluster emission spectrum within the cooling radius 
and the best-fit model are shown in Fig.~\ref{spec}. The cooling power within the 
cooling radius was estimated in 2-10 keV band and was found to be L$_{cool}$ = 
8.48$_{-0.28}^{+0.15}\,\times\,10^{43}\,$ erg\,s$^{-1}$\,.

\subsection{Peculiar features in the cluster}

\begin{figure*}
\centering
\includegraphics[width=8cm,height=7.5cm]{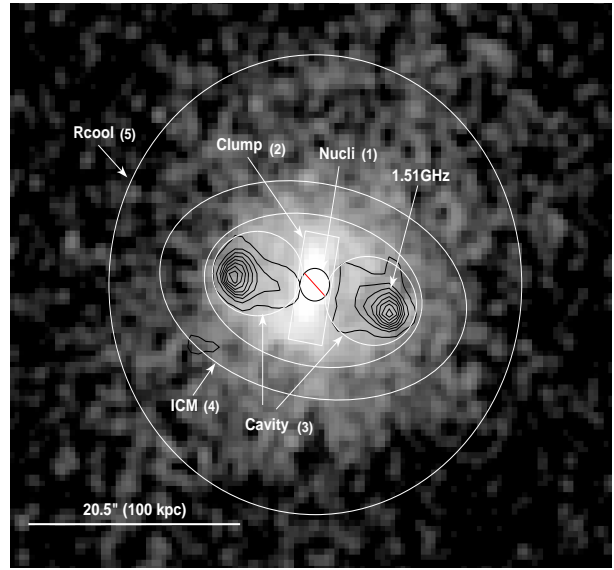} 
\caption{{\bf 2-sigma smoothed} 0.5--3 keV original {\it Chandra} image overlaid 
with the 1.51 GHz \textit{VLA} radio contours of 3C~320. Data from the selected 
regions highlighted in this figure were used for spectral investigation of the ICM.}
\label{reg} 
\end{figure*}

3C~320 galaxy shows various interesting features in its core region. For detailed 
investigation of these features, spectra were extracted from selective regions to 
examine the temperature and metal abundance distribution. The regions used for 
the selection are marked in Fig.~\ref{reg}. In this figure, the radio contours 
(black colour), generated from 1.51 GHz {\it NRAO/VLA Archive Survey} (Project 
ID: AG250), are overlaid on the 2$\sigma$ smoothed {\it Chandra} image. The 
lowest contour line of 1.51 GHz image is at 3 mJy/beam whereas the noise 
level (rms) is 80 $\mu$Jy/beam. These contours appear to fill the X-ray cavities 
along East and West directions. Spectra for each of the selected regions in the 
figure were extracted separately. To study the spectral properties of the central 
point source in 3C~320, spectrum was extracted from a circular region of 1.5 arcsec 
radius, centered on the peak of the X-ray emission (RA=15h 31m 25.38s; DEC=+35d 33m 
40.46s). The central source spectrum was fitted well with the APEC and power law 
model independently and the best fit parameters are given in Table~\ref{tab1}. 
Spectra, extracted from other marked regions in Fig.~\ref{reg} such as clumpy 
(excluding central nuclear) region marked as 2, cavity regions (marked as 3) 
and surrounding cavity (ICM) region (marked as 4) were also fitted with APEC 
model and the best-fit parameters are quoted in Table~\ref{tab1}.

\begin{table}
\setlength{\extrarowheight}{0.3cm}
\caption{Spectral parameters of selective regions in 3C~320. During spectral fitting, 
the metallicity was fixed at 0.5 Z$_{\odot}$.}
\vbox{%
\centering
\begin{tabular}{@{}lccccr@{}}
\hline
Region 		 & kT (keV) 			&$\Gamma$ 	& Z (Z$_{\odot}$) &  $\chi^2/dof$ \\
\hline
ICM       	 & 4.29$_{-0.40}^{+0.49}$ 	&--		& (0.50)		& 39.36$/$36 \\
East Cavity    	 & 3.74$_{-0.39}^{+0.36}$ 	&--		& (0.50)		& 36.12$/$35 \\ 
West Cavity    	 & 3.00$_{-0.32}^{+0.30}$ 	&--		& (0.50)		& 20.21$/$22 \\
Clumpy region    & 2.99$_{-0.30}^{+0.30}$ 	&--		& (0.50) & 40.83$/$35 \\
\multirow{2}{*}{Nucleus} & 3.27$_{-0.38}^{+0.56}$ 	&--	& (0.50) & 22.28$/$22 \\
                                            &  --	&2.14$_{-0.12}^{+0.13}$ &--   & 22.69$/$22 \\
\hline
\end{tabular}}
\label{tab1}
\end{table}
 
 \section{Discussion}
 
\subsection{Cavity energetics}

X-ray deficient regions (cavities) in the environment of galaxy clusters are known 
to be created due to the AGN outburst. Total outburst energy ($E_{cav}$) of an AGN can 
be estimated by measuring the volume `$V$' of the cavities and the pressure (`$p$') of 
surrounding ICM. Fig.~\ref{res} showed a pair of cavities in 3C~320 along the East 
and West directions associated with the central engine of the galaxy. These cavities 
are assumed to be ellipsoidal, symmetric about the plane of the sky, centres of which 
lie in the plane perpendicular to the line of sight \citep{2010ApJ...720.1066C}. Using 
these assumptions, physical parameters of each of the cavity were estimated. The size 
of the cavities, marked with white ellipses in Fig.~\ref{reg}, were determined through 
visual inspection as done in other cases \citep{2004ApJ...607..800B,2011ApJ...732...71C,
2016MNRAS.461.1885V}. Radio lobes due to the AGN jets are known to fill the X-ray deficient 
regions (cavities) in the ICM. The energy stored in each of the cavity was estimated by 
using the relation \citep{2004ApJ...607..800B,2006ApJ...652..216R}

\begin{equation}
E_{cav} = \frac{\gamma}{\gamma - 1} pV
\end{equation}

\begin{figure}
\centering
\includegraphics[width=8cm,height=8cm]{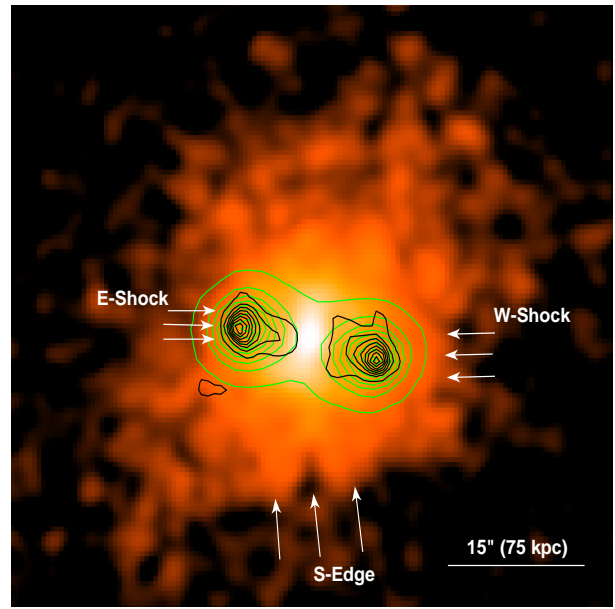} 
\caption{0.5--3 keV {\it Chandra} image of 3C~320, overlaid with 1.51 GHz (black) and 
1.55 GHz (green) ({\it VLA}) radio contours. Locations of shock and edge in the ICM of 
3C~320 galaxy are marked in the figure.}
\label{edge} 
\end{figure} 

where $V$ represents the volume of the cavity given by $V$ = 4$\pi R_l R^2_w$/3, 
$R_l$ and $R_w$ are the semi-major and semi-minor axes and $p$ is the thermal 
pressure exerted by radio bubbles on the surrounding ICM. Assuming that cavities 
are filled with the relativistic fluid, we chose $\gamma$=4/3. The cavity power 
($P_{cav}$) injected into the ICM was estimated by dividing the energy content of 
the cavity by its age (i.e. $P_{cav} = E_{cav}/t_{age}$), where cavity age was 
calculated by three different methods viz. buoyant rise time ($t_{buoy}$), refill 
time ($t_{refill}$) and sound crossing time ($t_{sonic}$) \citep{2000ApJ...534L.135M,2004ApJ...607..800B}. Here, $t_{buoy}$ is the time taken by 
the bubble to rise buoyantly to attain terminal velocity, $t_{refill}$ is the time 
to refill the volume which was displaced by the bubble and $t_{sonic}$ is the time 
required for the cavities to reach the present location from the centre, moving at 
the local sound speed. In the present study, we assume that $t_{buoy}$ provides 
good estimate of age of the cavity whereas $t_{refill}$ and $t_{sonic}$ provide 
lower and upper limits, respectively. Table~\ref{tab2} presents the physical 
parameters of cavities, ages and their mechanical energy contents.

As described above, average pressure surrounding the radio bubble was found to be 
$p$ = 1.08$\pm$0.31$\times$10$^{-10}$ erg cm$^{-3}$. Using this value, the total 
cavity energy (sum of E-cav and W-cav energies) was derived to be $E_{cav}$ = 
7.70$\pm$2.20$\times$10$^{59}$ erg. The AGN feedback power was calculated by using 
the values of E$_{cav}$ and t$_{age}$. In this calculation, we assume that the 
$t_{buoy}$ provides good estimate of the cavity age while $t_{refill}$ and $t_{sonic}$ 
gives the upper and lower limits, respectively. Here $t_{age}$ is taken to be the average 
age of E-cav and W-cav for all three estimations. Using these assumptions, the total 
AGN feedback power was estimated to be $P_{cav}$ = 3.52$_{1.85}^{6.75}\times$10$^{44}$ 
erg s$^{-1}$. The total mechanical power required to inflate the cavities is found to 
be approximately twice the cooling power, suggesting that the AGN hosted by this system 
delivers sufficient power to balance the radiative cooling loss of the ICM.

\begin{table}
\setlength{\extrarowheight}{0.3cm}
{\renewcommand{\arraystretch}{1.3} 
\caption{\label{tab2} Physical parameters of the cavities}
\begin{tabular}{@{}lcr@{}}
\hline
Parameters &  E-cavity  & W-cavity \\
\hline
Semi-major axis ($R_l$; in kpc) &21.3  &21.8 \\
Semi-minor axis ($R_w$; in kpc) &18.0  &18.7 \\
Projected distance from centre ($R$; in kpc)    &36.87 & 38.67 \\    
Volume (in $10^{68}\,$ cm$^{3}$) & 8.39 & 9.32 \\ 
E$_{cav}=4pV$ ($10^{59}\,$ erg) & 3.6$\pm$1.1 & 4.0$\pm$1.2\\
$t_{sonic}$ ($10^7\,$ yr) & 3.5$\pm$0.18 & 3.7$\pm$0.19\\
$t_{buoy}$ ($10^7\,$ yr) &  6.7$\pm$2.32 & 7.1$\pm$2.45 \\
$t_{refill}$ ($10^7\,$ yr) & 12.8$\pm$8.90 & 13.4$\pm$9.26 \\
\hline
\end{tabular}}
\footnotesize
\end{table}


\begin{figure*}
\vbox{
\includegraphics[scale=0.4]{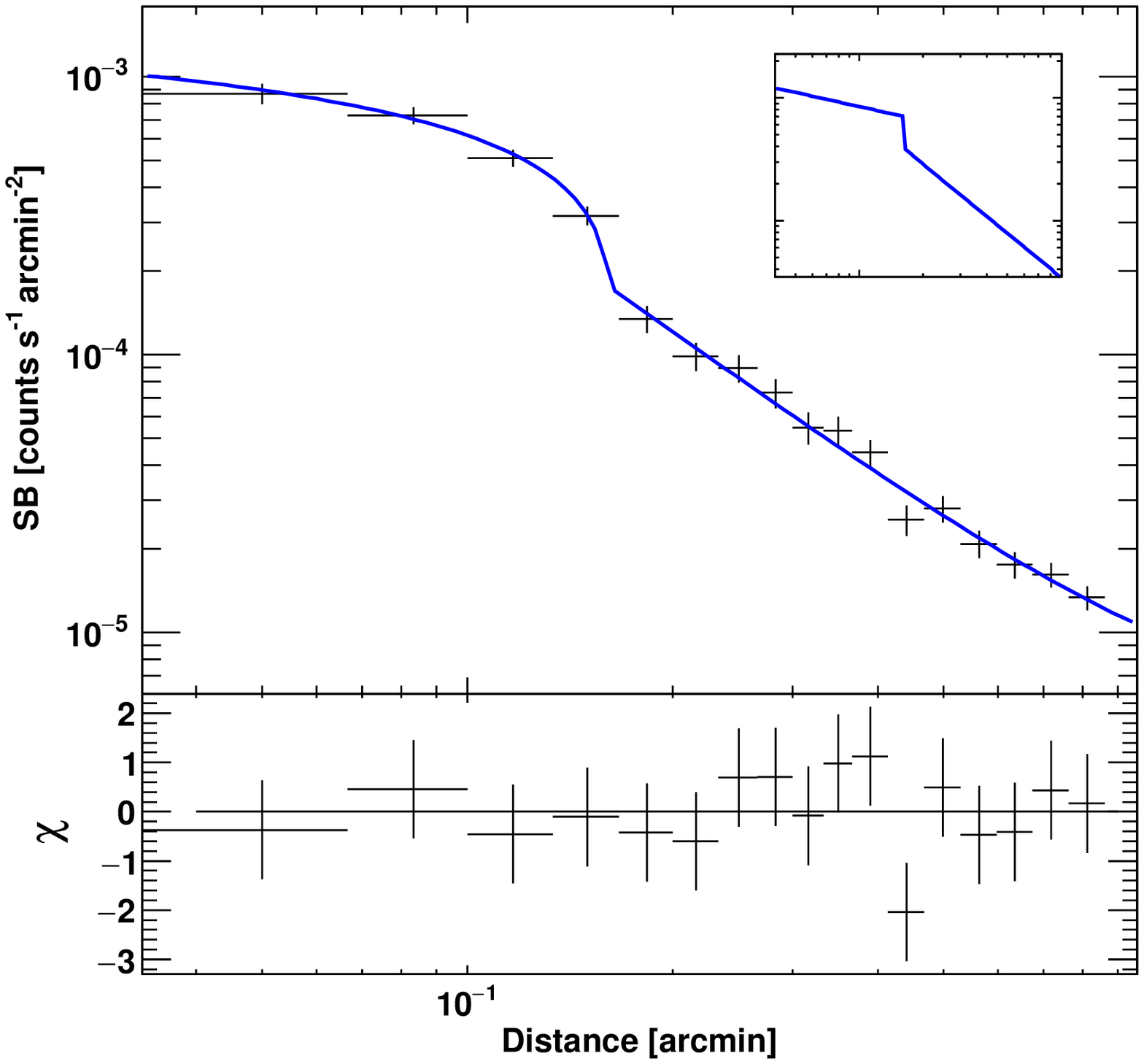}
\includegraphics[trim={2cm 0 0 0},scale=0.4]{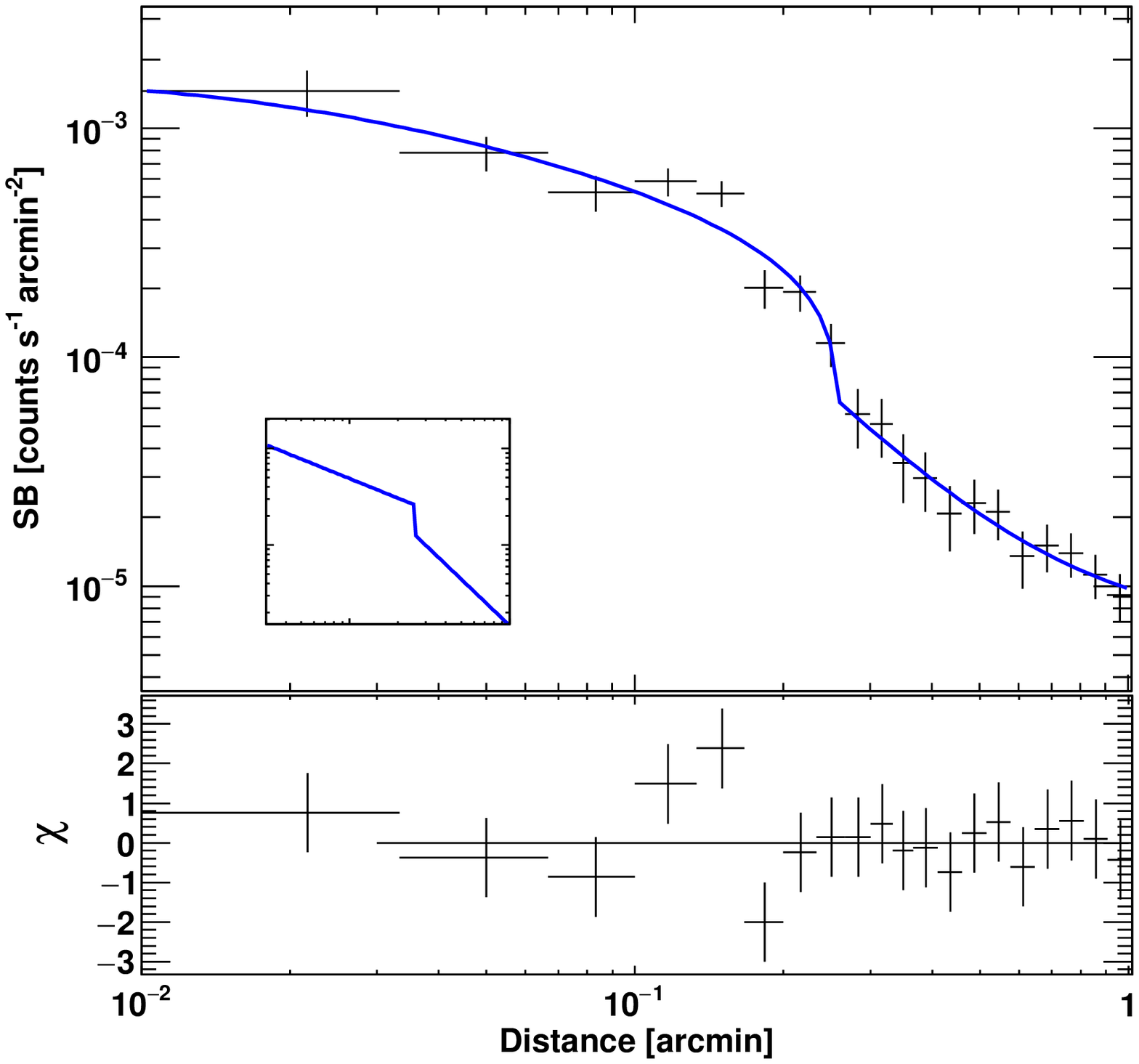}}
\caption{X-ray surface brightness profiles in 0.5--3 keV range, extracted from sectors 
in the East (110 -- 200 degrees -- left panel) and West (340 -- 400 degrees -- right 
panel) directions along with the the best-fit broken power-law density model. Inset in 
the figures show 3D simulated gas density model. The bottom panels of both the figures 
represent the residuals obtained from the fitting.}
\label{shock}
\end{figure*}

\subsection{Shock around bubbles}

Figure~\ref{edge} shows the 2$\sigma$ Gaussian smoothed 0.5--3 keV {\it Chandra} image 
overlaid with the 1.51 GHz in black (Project ID: AG250) and 1.55 GHz in green (Project 
ID: AG183) radio contours generated using NRAO/VLA Archive Survey of FR~II 3C~320. Here, 
the lowest contour level of 1.55 GHz map is of 7.7 mJy/beam at the noise level (rms) of 
25 $\mu$Jy/beam. This figure provides a hint to the presence of shocks in front of the 
radio lobes along the East and West directions due to their interaction with the ICM. 
Detection of shocks associated with AGN outbursts are very rare and are seen in a few 
cases in deep {\it Chandra} observations \citep{2005ApJ...628..629N,2011ApJ...732...13G,
2006MNRAS.366..417F,2009A&A...495..721S,2014MNRAS.442.3192V,2007ApJ...665.1057F,
2010MNRAS.407.2046M,2012ApJ...749...19K,2011ApJ...734L..28C,2017MNRAS.466.2054V}. 
To confirm the presence of shocks around the radio lobes in 3C~320, we generated 
0.5--3 keV surface brightness profiles along the directions of radio lobes i.e. 110--200 
degree and 340--400 degree angular regions in the East and West directions, respectively 
(Figure~\ref{shock}). The profiles were then fitted with the deprojected broken power law 
density model (bknpow) using {\it PROFFIT} (ver.1.4) package \citep{2011A&A...526A..79E}. 
This model assumes spherically symmetric hydrodynamic model for shock produced due to point 
explosion at the nucleus \citep{2005ApJ...625L...9N}. The best fit broken power law model 
along with the surface brightness profile towards the East and West are shown in 
Fig.~\ref{shock}. In our surface brightness profile fitting, the inner slope ($\alpha$1), 
outer slope ($\alpha$2), jump location (R$_{sh}$), density jump (C) and normalization 
(N$_0$) were kept free. The resultant best-fit parameters for the profiles along the East 
and West directions are given in the first and second rows of Table~\ref{tab3}. The density 
jump (C) obtained from fitting were then converted to the shock Mach numbers by using 
the adiabatic index $\gamma$= 5/3 and Rankine-Hugoniot jump condition for gas. The Mach 
numbers at corresponding jump locations were derived to be 1.5$\pm$0.38 and 1.8$\pm$0.66 
along East and West directions, respectively. For further confirmation of the shock 
features along these directions, we tried to fit the profiles by using broken beta (bknbeta) 
model. In this case, the inner part of the profile was fitted with the beta model while the 
outer part was treated with a power law component. The results obtained from this analysis 
are given in Table~\ref{tab3} and are in agreement with those obtained from the broken power 
law (bknpow) fit analysis. These results confirm the presence of shocks at $\sim$10$\arcsec$ 
(47 kpc) along the East and $\sim$16$\arcsec$ (76 kpc) along the West directions. The 
locations of the shocks and cavities are very close to one another as apparent in 
Fig.~\ref{edge}, implying that the origin of shocks and cavities are either due to an 
AGN outburst or multiple episodes of the AGN outbursts. However, the multiple episodes 
of AGN outburst in this case are unlikely to happen due to the missing evidences of 
multiple cavities. Therefore, a single AGN outburst as the cause of origin of close 
locations of shocks and cavities is more appropriate.

\begin{table*}
\setlength{\extrarowheight}{0.3cm}
\caption{Parameters of broken power law {\bf and broken beta}  density model}
\centering
\small
\begin{tabular}{cccccccc}
\hline
region-(model)  &	$\alpha$1/$\beta$ & r$_c$ &	$\alpha$2  & R$_{sh}$/R$_{fr}$ & C  & N$_0$ & $\chi^2/dof$  \\
   &         &    (arcmin)       &        &  (arcmin)   &     &      ($10^{-4}\, counts\, s^{-1} arcmin^{-2}$)  &    \\
\hline
E-shock-(bknpow)   & 0.38$\pm$0.16 & -- &	1.42$\pm$0.11 &	0.16$\pm$0.005	 &  1.80$\pm$0.19 &	35$\pm$6.90   &  13.72/13     \\
W-shock-(bknpow)  & 0.67$\pm$0.13 & -- &	1.58$\pm$0.53 &	0.26$\pm$0.020	 &  2.06$\pm$0.64 &	10.39$\pm$2.0 &  14.98/13  \\
{\bf S-edge}-(bknpow)  & 1.36$\pm$0.04 & -- & 1.31$\pm$0.04 &	0.28$\pm$0.016	 &  1.60$\pm$0.06 &	3.5$\pm$0.14  &  11.04/9  \\
E-shock-(bknbeta)  & 0.23$\pm$0.03 &	0.013$\pm$0.009 &	1.52$\pm$0.12	 &  0.17$\pm$0.002 &	1.57$\pm$0.18 & 880$\pm$620   &  8.37/7  \\
W-shock-(bknbeta)  & 0.23$\pm$0.06 &	0.03$\pm$0.01 &	1.73$\pm$0.4	 &  0.24$\pm$0.01 &	1.83$\pm$0.47 & 288$\pm$86 &  17.2/7  \\
{\bf S-edge}-(bknbeta)  & 0.49$\pm$0.003 &	0.05$\pm$0.005 &	0.51$\pm$0.004	 &  0.28$\pm$0.004 &	1.78$\pm$0.03 & 616$\pm$19  &  6.56/7 \\
\hline
\end{tabular}\label{tab3}
\end{table*}

\subsection{Cold fronts}

\begin{figure}
\includegraphics[scale=0.4]{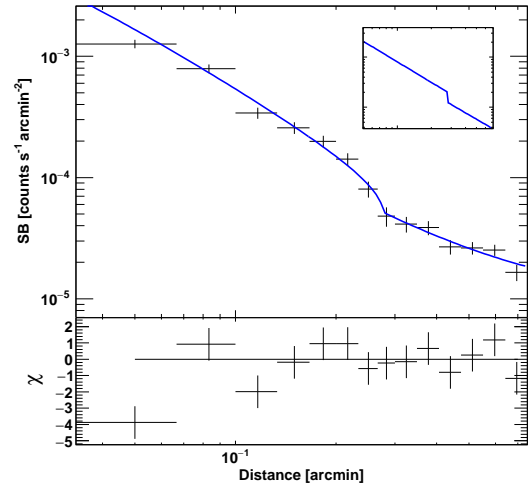}
\caption{X-ray surface brightness profile in 0.5--3 keV band extracted from sector 
along the South (250\degr -- 310\degr) direction. The profile was fitted with broken 
power law density model and is shown along with the best-fit model in solid line. 
Inset in the figure shows 3D simulated gas density model. The bottom panel exhibits 
the residuals obtained from the fit.}
\label{front}
\end{figure}

Sharp surface brightness edges (discontinuities in the surface brightness profile) 
forming arc-like structures in cool core clusters represent signatures of recent 
mergers \citep{2000ApJ...534L.135M,2007PhR...443....1M,2009ApJ...704.1349O}. 
Fig.~\ref{edge} also revealed a surface brightness edge along the South (S) 
direction at radius of $\sim$16\arcs (82 kpc). To investigate this feature in 
detail, we extracted surface brightness profile in the direction of the edge e.g. 
S (250--310 degree sectoral region) which was then fitted with the deprojected 
broken power law (Fig.~\ref{front}) and broken beta density model independently. 
This surface brightness profile confirmed a sharp density jump at $\sim$82 kpc 
and was found to be of $\sim$1.6--1.7. The best-fit parameters obtained from the 
broken power law and broken beta model fit are summarized in Table~\ref{tab3}. 
Azimuthally averaged temperature profile (Figure~\ref{therm}) also showed temperature 
jump at approximately the same location (the small variation in radius may be due to 
5\arcsec\, binning). Due to poor statistics, spectral fitting to identify the 
temperature jumps along the edge direction is difficult. These features (detection 
of edge, density jump and temperature jump) are known to be associated with the 
contact discontinuities at the edge of gas clouds that are rapidly moving through 
the ambient gas (less dense and hot medium) called as ``cold fronts'' 
\citep{2000ApJ...541..542M}. Thus, the analysis indicates the possible 
detection of a cold front in the ICM of 3C~320. Such detections are very rare 
in the ICM in the systems at moderate redshifts. The possible detection of the
cold front in the ICM of 3C~320 point towards the ongoing merger activity 
\citep{2009ApJ...704.1349O}. 
 
A careful search of literature revealed that the highest redshift cluster hosting such a 
cold front is MACS~J1149.6+223 ($z$=0.54; \citealt{2016ApJ...819..113O}), while all 
other detections are at low or moderate redshifts ($z <$0.2), 3C~320 ($z$=0.342; 
present work), Abell~2744 ($z$=0.308; \citealt{2011ApJ...728...27O}), 1ES0657-558 
($z$=0.296; \citealt{2009ApJ...704.1349O}), Abell~521 ($z$=0.25; 
\citealt{2013ApJ...764...82B}), 4C+55.16 ($z$=0.2412; \citealt{2011MNRAS.415.3520H}), 
Abell~2163 ($z$=0.201; \citealt{2001ApJ...563...95M}). The non-detection of cold 
fronts at high redshifts is probably due to the lack of availability of deep 
high spatial resolution detectors till date.

\subsection{Cooling and Star Formation}

This cluster experiences radiative cooling in its core with cooling time 
$\sim$7 $\times$ 10$^8$ yr (Fig.~\ref{tcool}) and cooling rate $\dot{M}$ = 
192$\pm$3 M$_{\odot}$ yr$^{-1}$ ($\dot{M}=2L\mu m_p/5kT$; \citealt{2012Natur.488..349M}). 
In 3C~320, the observed radio jets originating from the AGN feedback results in 
the formation of bubbles/cavities in the ICM. As the bubbles inflate, the cool, low 
entropy, metal-rich gas in the ICM is pushed outward forming the plume-like features 
around the bubbles/cavities \citep{2011MNRAS.415.3520H}. It is also believed 
that these plume-like features around the bubbles/cavities can trigger star 
formation in the ICM. Therefore, we tried to estimate the star formation rate 
(SFR) by using the far-infrared 60 $\micron$ luminosity of the cluster 6.67 $\times$ 
10$^{44}$ erg s$^{-1}$ (\citealt{2002A&A...381..389A}). The far-infrared 
luminosity was estimated by using the relation L$_{FIR}$ = 1.7 L$_{60 \micron}$ 
\citep{1997MNRAS.289..490R}. Using this luminosity and the relation given by \cite{2006ApJ...652..216R}, the SFR is given as, 
\begin{equation}
\frac{SFR}{M_{\odot}\, yr^{-1}} \leq 4.5 \left(\frac{L_{FIR}}{10^{44}\, erg\, s^{-1}}\right)
\end{equation}

 This estimate provides the upper limit of star formation rate in 3C~320 as $\sim$51 
M$_{\odot}$ yr$^{-1}$. \cite{2015ApJ...811...73L} during the study of cool core clusters, 
estimate the SFR over a wide range (from 0 to few 100 M$_{\odot}$ yr$^{-1}$) with an average of 
40 M$_{\odot}$ yr$^{-1}$. The measured SFR in the ICM of this cluster is in agreement with 
\cite{2015ApJ...811...73L} and is found to be approximately 1/4$^{th}$ of the cooling rate. 

\section{Conclusions}
We presented results obtained from the systematic analysis of a total of 110 ks 
{\it Chandra} observations of a moderate redshift ($z$=0.342) galaxy cluster 3C~320. 
A pair of prominent X-ray cavities at an average distance of $\sim$38 kpc from its centre 
are detected along the East and West directions of 3C~320. The total outburst energy, 
age and cavity power were estimated to be $\sim$7.70 $\times$ 10$^{59}$ erg, $\sim$7 
$\times$ 10$^7$ yr and $\sim$3.5 $\times$ 10$^{44}$ erg s$^{-1}$, respectively. The 
cooling luminosity within the cooling radius of $\sim$100 kpc was found to be $L_{cool} 
\sim 8.5\, \times 10^{43}$ erg s$^{-1}$, lower than the cavity power. This indicates 
that the cavity power is sufficient enough to balance the ICM cooling. AGN driven weak 
shocks at $\sim$47 kpc and $\sim$76 kpc from the cluster centre along the East and West 
directions, respectively, were also detected around the radio bubbles. Using the observed 
density jumps of $\sim$1.8 and $\sim$2.1 at the shock locations along the East and West 
directions, respectively, the Mach numbers were yielded to be $\sim$1.6 and $\sim$1.8. 
A surface brightness edge was also detected in the {\it Chandra} image along the South 
direction at $\sim$82 kpc from the centre. The density jump at the edge location was 
estimated to be $\sim$1.6 and is probably due to the presence of a cold font. 
3C~320 is the second highest redshift cluster that hosts cold front. Using the FIR 
luminosity, the star formation rate was estimated to be $\sim$ 51 M$_{\odot} yr^{-1}$ 
which is 1/4$^{th}$ of the cooling rate.

\section{Acknowledgments} 
The authors sincerely thank the anonymous referee for constructive comments and 
valuable suggestions on the paper. This work has made use of data from the \textit{Chandra}, 
\textit{VLA} and Gemini-South archive, NASA's Astrophysics Data System(ADS), 
Extragalactic Database (NED), software provided by the \textit{Chandra} X-ray 
Centre (CXC), HEASOFT for spectral fitting and Veusz plotting tools. NDV thanks 
to Science and Engineering Research Board (SERB), India for providing research 
fund (Ref.No.: YSS/2015/001413). NDV also thanks to S. Sonkamble for  his help 
in obtaining few of the plots in this paper and also for scientific discussion.

\def\aj{AJ}%
\def\actaa{Acta Astron.}%
\def\araa{ARA\&A}%
\def\apj{ApJ}%
\def\apjl{ApJ}%
\def\apjs{ApJS}%
\def\ao{Appl.~Opt.}%
\def\apss{Ap\&SS}%
\def\aap{A\&A}%
\def\aapr{A\&A~Rev.}%
\def\aaps{A\&AS}%
\def\azh{AZh}%
\def\baas{BAAS}%
\def\bac{Bull. astr. Inst. Czechosl.}%
\def\caa{Chinese Astron. Astrophys.}%
\def\cjaa{Chinese J. Astron. Astrophys.}%
\def\icarus{Icarus}%
\def\jcap{J. Cosmology Astropart. Phys.}%
\def\jrasc{JRASC}%
\def\mnras{MNRAS}%
\def\memras{MmRAS}%
\def\na{New A}%
\def\nar{New A Rev.}%
\def\pasa{PASA}%
\def\pra{Phys.~Rev.~A}%
\def\prb{Phys.~Rev.~B}%
\def\prc{Phys.~Rev.~C}%
\def\prd{Phys.~Rev.~D}%
\def\pre{Phys.~Rev.~E}%
\def\prl{Phys.~Rev.~Lett.}%
\def\pasp{PASP}%
\def\pasj{PASJ}%
\def\qjras{QJRAS}%
\def\rmxaa{Rev. Mexicana Astron. Astrofis.}%
\def\skytel{S\&T}%
\def\solphys{Sol.~Phys.}%
\def\sovast{Soviet~Ast.}%
\def\ssr{Space~Sci.~Rev.}%
\def\zap{ZAp}%
\def\nat{Nature}%
\def\iaucirc{IAU~Circ.}%
\def\aplett{Astrophys.~Lett.}%
\def\apspr{Astrophys.~Space~Phys.~Res.}%
\def\bain{Bull.~Astron.~Inst.~Netherlands}%
\def\fcp{Fund.~Cosmic~Phys.}%
\def\gca{Geochim.~Cosmochim.~Acta}%
\def\grl{Geophys.~Res.~Lett.}%
\def\jcp{J.~Chem.~Phys.}%
\def\jgr{J.~Geophys.~Res.}%
\def\jqsrt{J.~Quant.~Spec.~Radiat.~Transf.}%
\def\memsai{Mem.~Soc.~Astron.~Italiana}%
\def\nphysa{Nucl.~Phys.~A}%
\def\physrep{Phys.~Rep.}%
\def\physscr{Phys.~Scr}%
\def\planss{Planet.~Space~Sci.}%
\def\procspie{Proc.~SPIE}%
\let\astap=\aap
\let\apjlett=\apjl
\let\apjsupp=\apjs
\let\applopt=\ao
\bibliographystyle{mn2e}
\bibliography{mybib}




\end{document}